\documentclass[aps,prb,floats,twocolumn,amsmath,amssymb,superscriptaddress]{revtex4-2}
\usepackage[utf8]{inputenc}
\usepackage{graphicx}
\usepackage{epsfig}
\usepackage{amsfonts}
\usepackage{amsmath}
\usepackage{natbib}
\usepackage{multirow}
\usepackage{bm}
\usepackage{epstopdf}
\usepackage{ulem}
\usepackage{color}
\usepackage{hyperref}
\usepackage[utf8]{inputenc}
\usepackage{color}
\usepackage{listings}
\usepackage{physics}

\definecolor{green}{rgb}{0.0, 0.72, 0.92}

\usepackage{caption}
\usepackage{subcaption}
\definecolor{gray}{rgb}{0.4,0.4,0.4}
\definecolor{darkblue}{rgb}{0.0,0.0,0.6}
\definecolor{cyan}{rgb}{0.0,0.6,0.6}

\usepackage{xcolor}

\newcommand{\exciting}{{\usefont{T1}{lmtt}{b}{n}exciting}}
\newcommand{\ie}{{\it i.e.}, }
\newcommand{\eg}{{\it e.g.}, }

\lstdefinelanguage{XML}
{
	morestring=[b]",
	morestring=[s]{>}{<},
	morecomment=[s]{<?}{?>},
	stringstyle=\color{black},
	identifierstyle=\color{blue},
	keywordstyle=\color{cyan},
	morekeywords={xmlns,version,type}
}

\begin{document}
	\title{Accurate and efficient treatment of spin-orbit coupling via second variation employing local orbitals}
	\author{Cecilia Vona}
	\thanks{These two authors contributed equally.}
	\author{Sven Lubeck}
	\thanks{These two authors contributed equally.}
	\author{Hannah Kleine}
	\affiliation{Institut f\"{u}r Physik and IRIS Adlershof, Humboldt-Universit\"{a}t zu Berlin, 12489 Berlin, Germany}
	\author{Andris Gulans}
	\affiliation{Department of Physics, University of Latvia, LV-1004 Riga, Latvia}
	\author{Claudia Draxl}
	\affiliation{Institut f\"{u}r Physik and IRIS Adlershof, Humboldt-Universit\"{a}t zu Berlin, 12489 Berlin, Germany}
	\affiliation{European Theoretical Spectroscopic Facility (ETSF)}
	\date{\today}
	\begin{abstract}
		A new method is presented that allows for efficient evaluation of spin-orbit coupling (SOC) in density-functional theory calculations. In the so-called second-variational scheme, where Kohn-Sham functions obtained in a scalar-relativistic calculation are employed as a new basis for the spin-orbit-coupled problem, we introduce a rich set of local orbitals as additional basis functions. Also relativistic local orbitals can be used. The method is implemented in the all-electron full-potential code \exciting. We show that, for materials with strong SOC effects, this approach can reduce the overall basis-set size and thus computational costs tremendously. 
	\end{abstract}
	\maketitle
	
	\section{Introduction}
	Spin-orbit coupling (SOC) is crucially important for accurate electronic-structure calculations of many materials. To illustrate, SOC is responsible for lifting the degeneracy of low-energy excitons in transition-metal dichalcogenides (TMDCs) \cite{Marsili_2021,Diana_2016,Caruso_2022}, opening a tiny gap in graphene \cite{Kane_2005,Yao_2007,Gmitra2009}, and dramatically lowering the fundamental band gap in halide perovskites \cite{Even_2013}. However, the impact of SOC is not limited to changing features of the electronic bands. It affects bond lengths \cite{vanLenthe1996,Oliveira_2013,gulans_2022}, phonon energies \cite{vanLenthe1996,DalCorso_2013}, and even turns deep defects into shallow ones~\cite{Ming2022}.
	
	In density-functional-theory (DFT) computations, SOC is treated differently in the various methods and codes. In this context, the family of full-potential linearized augmented planewaves (LAPW) methods is commonly used as the reference, \eg  for new implementations \cite{huhn_2017} or for assessing pseudopotentials \cite{Schlipf2015}. Commonly used LAPW codes \cite{Gulans_2014,Elk,Blugel2006,Blaha1990} employ similar strategies to account for SOC. For the low-lying core orbitals, the standard approach is to solve the radial 4-component Dirac equation, assuming a spherically symmetric potential. For the semi-core and valence electrons, the common strategy is to employ a two-step procedure \cite{singh_1994}. First, the Kohn-Sham (KS) problem is solved within the scalar-relativistic approximation (first variation, FV). Then, the solutions of the full problem including SOC are constructed using the FV wavefunctions as the basis. This step is known as the second variation (SV). The strategy relies on an assumption that SOC introduces a small perturbation, and, indeed, this scheme is appropriate and efficient for many materials, since all the occupied and only a handful of unoccupied bands are sufficient. Under these circumstances, the two-step procedure offers a clear computational advantage over methods where SOC is treated on the same footing with other terms of the Hamiltonian~\cite{Loucks1965,Kutepov2021,gulans_2022}.
	
	Some materials, however, require more involved calculations than others. For example, it was argued by Scheidemantel and coworkers \cite{Scheidemantel2003} that Bi$_2$Te$_3$ requires the consideration of unoccupied bands of at least 8 Ry above the Fermi level to give reliable results. Even more striking, in the halide perovskites, the full set of KS orbitals is needed for convergence~\cite{Vona_2022}. These cases illustrate that for some materials SOC cannot be considered as a small perturbation. Moreover, it is known that scalar- and fully-relativistic orbitals have different asymptotic behavior at small distances from the nuclei. Most notably, SOC introduces a splitting within the $p$-orbitals into spinors, where the  radial part of the $p_{3/2}$ solution goes to zero, while the $p_{1/2}$ one diverges. This behavior cannot be recovered in terms of scalar-relativistic (SR) functions. Therefore, in Refs.~\onlinecite{singh_1994} and \onlinecite{Kunes2001}, the SV basis was extended by additional basis functions, local orbitals (LOs), that recover the correct asymptotic behavior of the $p_{1/2}$ orbitals. This approach is a step forward compared to conventional SV calculations. By introducing, however, exactly one shell of $p_{1/2}$ LOs per atom, it does not offer the possibility of systematic improvement toward the complete-basis-set limit. There are examples of SR calculations in literature where an extensive use of LOs is required to reach precision targets \cite{Michalicek2013, Nabok2016, Zavickis2022}. Furthermore, Ref.~\onlinecite{Kutepov2021} demonstrated this point also in the context of fully-relativistic calculations. We therefore conclude that the state-of-the-art SV approaches, be it with or without $p_{1/2}$ LOs, are not sufficient for a systematic description of SOC in condensed-matter systems. 
	
	In this work, we introduce a new approach, termed \textit{second variation with local orbitals} (SVLO), which makes use of the fact that relativistic effects are strongest around the atomic nuclei. Therefore, in comparison to the standard SV approach, it is important to increase the flexibility of the basis specifically in these regions. To satisfy this need, we express the solution of the full problem in terms of FV wavefunctions and rich sets of LOs. In contrast to the usual approach, all LOs are treated as explicit basis functions, also on the SV level. In addition, we include LOs obtained from solving the Dirac equation (termed Dirac-type LOs in the following) beyond $p_{1/2}$ functions. Based on the implementation in the all-electron full-potential package \exciting~\cite{Gulans_2014}, we demonstrate and validate our method in band-structure and total-energy calculations of Xe, MoS$_2$, PbI$_2$, $\gamma$-CsPbI$_3$, and Bi$_2$Te$_3$.
	
	\section{Method}
	\subsection{Conventional second variation}
	\label{sec:SV_method}
	We consider the two-component KS equations
	\begin{equation}\label{eqn:2cKSeqn}
		\sum_{\sigma' = \,\uparrow, \downarrow} \hat{H}_{\sigma \sigma'} \Psi_{i \bm{k}\sigma'}\left( \bm{r} \right) = \varepsilon_{i \bm{k}} \Psi_{i \bm{k}\sigma}\left( \bm{r} \right)
	\end{equation}
	for the spin components $\sigma =\:\uparrow, \downarrow$. The resulting single-particle spinors $\sum_\sigma \Psi_{i \bm{k}\sigma}\left( \bm{r} \right) \ket{\sigma}$ have eigenenergies $\varepsilon_{i \bm{k}}$, where $i$ is the band index and $\bm{k}$ the Bloch wave vector. The Hamiltonian $\hat{H}_{\sigma \sigma'}$ consists of a SR part and a spin-orbit part that couples the two spin components:
	\begin{equation}\label{eqn:SR-SO-split}
		\hat{H}_{\sigma \sigma'} = \delta_{\sigma \sigma'} \hat{H}^\text{SR}_\sigma + \hat{H}^\text{SOC}_{\sigma \sigma'} .
	\end{equation}
	As described in Refs.~\onlinecite{Loucks1965,Kutepov2021,gulans_2022}, Eq.~\ref{eqn:2cKSeqn} can be solved directly, \ie non-perturbatively (NP), requiring a significantly larger computational effort compared to a SR calculation. Given that $\hat{H}^\text{SOC}_{\sigma \sigma'}$ typically leads to a small correction, it is unsatisfactory to pay the full price for the NP solution. For this reason, often the conventional SV method is employed. In this approach, first, one solves the scalar-relativistic problem in FV,
	\begin{equation}\label{eqn:FV}
		\hat{H}^\text{SR}_\sigma \Psi^\text{FV}_{j \bm{k}\sigma}\left( \bm{r} \right) = \varepsilon^\text{FV}_{j \bm{k}\sigma} \Psi^\text{FV}_{j \bm{k}\sigma}\left( \bm{r} \right) ,
	\end{equation}
	and subsequently uses the resulting FV eigenstates (which are the FV KS wavefunctions) as a basis for the SV eigenstates
	\begin{equation}\label{eqn:SVbasis}
		\Psi^{\text{SV}}_{i \bm{k}\sigma}\left( \bm{r} \right) = \sum_{j} C^{\text{SV}}_{\bm{k}\sigma j i} \Psi^{\text{FV}}_{j \bm{k}\sigma}\left( \bm{r} \right) .
	\end{equation}
	Here, $j$ runs over all $N_\text{occ}$ occupied and a limited number $N_\text{unocc}$ of unoccupied FV KS states. Approximating the exact solution $\Psi_{i \bm{k}\sigma}\left( \bm{r} \right)$ by $\Psi^{\text{SV}}_{i \bm{k}\sigma}\left( \bm{r} \right)$, one obtains the SV eigenequation for the expansion coefficients $C^{\text{SV}}_{\bm{k}\sigma j i}$
	\begin{equation}\label{eqn:SVeigenequation}
		\sum_{\sigma' j'} H_{\bm{k}\sigma \sigma' j j'} C^{\text{SV}}_{\bm{k}\sigma' j' i } = \varepsilon^\text{SV}_{i \bm{k}} C^{\text{SV}}_{\bm{k}\sigma j i} ,
	\end{equation}
	where $H_{\bm{k}\sigma \sigma' j j'}$ are the matrix elements of $\hat{H}_{\sigma \sigma'}$, as defined in Eq.~\ref{eqn:SR-SO-split}, with respect to the basis functions $\Psi^{\text{FV}}_{j \bm{k}\sigma}\left( \bm{r} \right)$,
	\begin{eqnarray}
		H_{\bm{k}\sigma \sigma' j j'} &=& \bra{\Psi^{\text{FV}}_{j \bm{k}\sigma}} \hat{H}_{\sigma \sigma'} \ket{\Psi^{\text{FV}}_{j' \bm{k}\sigma'}} \nonumber \\
		&=& \delta_{\sigma \sigma'} \delta_{j j'} \varepsilon^\text{FV}_{j \bm{k}\sigma} + \bra{\Psi^{\text{FV}}_{j \bm{k}\sigma}} \hat{H}^\text{SOC}_{\sigma \sigma'} \ket{\Psi^{\text{FV}}_{j' \bm{k}\sigma'}}.
	\end{eqnarray}
	
	\subsection{Second variation with local orbitals}
	\label{sec:SVLO_method}
	
	The SVLO approach makes use of the underlying LAPW+LO method that is utilized to solve the FV problem in Eq.~\ref{eqn:FV}. Within the LAPW+LO method, KS orbitals are represented by two distinct types of basis functions, namely LAPWs, $\phi_{\bm{G} \bm{k}}\left( \bm{r} \right)$, and LOs, $\phi_{\mu}\left( \bm{r} \right)$, which are indexed by reciprocal lattice vectors $\bm{G}$ and LO indices $\mu$, respectively,
	\begin{equation}\label{eqn:LAPW+LObasis}
		\Psi^{\text{FV}}_{j \bm{k}\sigma}\left( \bm{r} \right) = \sum_{\bm{G}} C_{\bm{k}\sigma \bm{G} j} \phi_{\bm{G} \bm{k}}\left( \bm{r} \right) + \sum_{\mu} C_{\bm{k}\sigma\mu j} \phi_{\mu}\left( \bm{r} \right).
	\end{equation}
	In order to avoid linear dependency issues between FV eigenfunctions and LOs in our new approach, we modify these FV eigenfunctions such that LO contributions are neglected, and only the first sum in Eq. \ref{eqn:LAPW+LObasis} is further considered:
	\begin{equation}\label{eqn:FVbar}
		\bar{\Psi}^\text{FV}_{j \bm{k}\sigma}\left( \bm{r} \right) = \sum_{\bm{G}} C_{\bm{k}\sigma \bm{G} j} \phi_{\bm{G} \bm{k}}\left( \bm{r} \right) .
	\end{equation}
	We combine these modified FV functions with the original set of LOs to form the SVLO basis
	\begin{eqnarray}\label{eqn:SVLObasis}
		\Psi^{\text{SVLO}}_{i \bm{k}\sigma}\left( \bm{r} \right) =&& \sum_{j} C^{\text{SVLO}}_{\bm{k}\sigma j i} \bar{\Psi}^{\text{FV}}_{j \bm{k}\sigma}\left( \bm{r} \right) \nonumber\\  &+&\sum_{\mu} C^{\text{SVLO}}_{\bm{k} \sigma \mu i} \phi_{\mu}\left( \bm{r} \right).
	\end{eqnarray}
	The total basis-set size in the SVLO method includes the number of these LO basis functions, $N_\text{LO}$. To summarize, the total number of basis functions in the two methods is 
	\begin{equation}\label{eqn:N_b_SV_LO}
		N_\text{b}^{\mathrm{SV(LO)}} = \begin{cases}
			N_\text{occ} + N_\text{unocc} & \, \text{SV} \\
			N_\text{occ} + N_\text{unocc} + N_\text{LO} & \, \text{SVLO} .
		\end{cases}
	\end{equation}
	In both cases, $N_\text{unocc}$ is a computational parameter, and the results need to be converged with respect to it. We note in passing that the SVLO basis is not orthogonal and thus carries a slight computational overhead compared to the conventional SV method, since it leads to a generalized eigenvalue problem. 
	
	The SVLO method is implemented in \exciting. How the different types of LOs are constructed will be described in the next section. Unlike Ref.~\onlinecite{singh_1994,Kunes2001}, our approach uses the entire set of LOs from the FV basis (including Dirac-type LOs if necessary) as the basis in the SV step.
	
	\begin{table}[hb]
		\centering
		\caption{Structural information and convergence parameters used in the calculations of the considered  materials. $R_{\mathrm{MT}}^{\text{min}}G_{\mathrm{max}}$ is the product of the largest reciprocal lattice vector, $G_{\mathrm{max}}$, considered in the LAPW basis and the (smallest) MT radius, $R_{\mathrm{MT}}^{\mathrm{min}}$. For MoS$_2$, the latter refers to the S sphere ($R_{\mathrm{MT}}$=2.05 $a_0$). For more detailed information, we refer to the input files provided at NOMAD.}
		\begin{tabular}{l|c|c|c|c|c} 
			Material & Xe & MoS$_2$ & PbI$_2$ & CsPbI$_3$ & Bi$_2$Te$_3$ \\
			\hline
			$a$ [\AA] & 6.20 & 3.16  & 4.56 & 8.86 & 10.44 \\
			$b$ [\AA] & 6.20 & 3.16  & 4.56 & 8.57 & 10.44 \\
			$c$ [\AA] & 6.20 & 15.88  & 6.99 & 12.47 & 10.44\\
			Space group & Fm3-m &P-6m2& P-3m1& Pnam &R-3m\\
			Ref. & \onlinecite{wyckoff_book_1963},  \onlinecite{ashcroft_book_2011} & \onlinecite{Broeker_2001} & 
			\onlinecite{Palosz_1990} & \onlinecite{Sutton_2018}& \onlinecite{Akshay_2017}\\      \hline
			$R_{\mathrm{MT}}$ [$a_0$]&3.00 & 2.30/2.05 & 2.90 & 2.90 & 2.80 \\
			$R_{\mathrm{MT}}^{\text{min}}G_{\mathrm{max}}$& 8 & 8 & 8 & 9 & 10 \\
			$\bm{k}$-mesh & 4$\times$4$\times$4 & 6$\times$6$\times$1& 6$\times$6$\times$4 & 3$\times$3$\times$2 & 6$\times$6$\times$6  \\
		\end{tabular}
		\label{tab:conv_parameter}
	\end{table}
	
	\subsection{Local orbitals}
	LOs are basis functions with the characteristic of being non-zero only in a sphere centered at a specific nucleus~$\alpha$~\cite{singh_1991}. They take the form of atomic-like orbitals which read
	\begin{equation}
		\phi_{\mu}(\bm{r}) = \delta_{\alpha,\alpha_\mu}\delta_{l,l_\mu}\delta_{m,m_\mu} U_{\mu}(r_\alpha)Y_{lm}(\hat r_\alpha),
	\end{equation}
	where $Y_{lm}(\hat r_\alpha)$ are spherical harmonics and $U_{\mu}(r_\alpha)$ are linear combinations of two or more radial functions
	\begin{equation}\label{eqn:LO_SR_def}
		U^{\text{SR}}_{\mu}(r_\alpha) = \sum_{\xi}\! a_{\mu\xi}u_{\alpha\xi l}(r_\alpha;\varepsilon_{\alpha\xi l}).
	\end{equation}
	The index $\xi$ sums over different radial functions. These radial functions $u_{\alpha\xi l}(r_\alpha;\varepsilon_{\alpha \xi l})$ are the solutions of the SR Schr\"odinger equation and/or their energy derivatives (of any order), evaluated at predefined energy parameters $\varepsilon_{\alpha \xi l}$. Depending on their purpose all radial functions have the same energy parameter or one corresponding to a different state. To account for the asymptotic behavior of relativistic orbitals at the atomic nuclei, we build LOs in which the radial functions are solutions of the Dirac equation
	\begin{equation}
		U^{\text{Dirac}}_{\mu}(r_\alpha) =
		\sum_{\xi, J}\! a_{\mu \xi J}u_{\alpha\xi J l}
		(r_\alpha;\varepsilon_{\alpha\xi J l}).
	\end{equation}
	Here, the radial functions and the energy parameters are characterized by the additional quantum number $J$. We sum over the index $J$ to show that it is possible to combine radial functions with different total angular momentum (but the same angular momentum $l$). It is also possible to combine $J$-resolved radial functions with SR radial functions. In the following we call any LOs including at least one $J$-resolved radial function, Dirac-type LOs. With this, we can add one or more LOs with any relativistic quantum number, going beyond what has been suggested by Singh \cite{singh_1994}. This approach, also used in Ref.~\onlinecite{gulans_2022}, is convenient since the general form of the LOs of Eq.~\ref{eqn:LO_SR_def} is kept.

	\section{Computational details}
	We consider a set of five materials, including 3D and 2D semiconductors and a topological insulator, with different atomic species, stoichiometry, and degree of SOC. For all of them, we employ experimental atomic structures. All calculations are performed with the package \texttt{exciting}~\cite{Gulans_2014} where the new method is implemented. Exchange and correlation effects are treated by the PBE parametrization of the generalized gradient approximation \cite{Perdew_1996,Perdew_1997_e}. Core electrons are described by means of the 4-component Dirac equation considering only the spherically symmetric part of the KS potential. For semicore and valence electrons, the zero-order regular approximation \cite{Lenthe1993,Lenthe1994} is used to obtain the SR and SOC contributions to the kinetic energy operator. The SOC term is applied only within the muffin-tin region and is evaluated by the following expression:
	\begin{equation}\label{eqn:SOC-term}
		\hat{H}^\text{SOC} = 
		\frac{c^2}{(2c^2-V)^2}\frac{1}{r}\frac{dV}{dr}\boldsymbol{\sigma L},
	\end{equation}
	where $\boldsymbol{\sigma}$ and $V$ are the vector of Pauli matrices and the spherically symmetric component of the KS potential, respectively. 
	
	The structural and computational parameters are displayed in Table~\ref{tab:conv_parameter}. The respective $\bm{k}$-mesh and the dimensionless LAPW cutoff $R_{\mathrm{MT}}^{\mathrm{min}}G_{\mathrm{max}}$ are chosen such that total energies per atom and band gaps are within a numerical precision of $10^{-2}$ eV/atom and $10^{-2}$ eV, respectively. The actual LAPW basis cutoff $G_{\mathrm{max}}$ is determined by dividing $R_{\mathrm{MT}}^{\mathrm{min}}G_{\mathrm{max}}$ by the smallest MT radius $R_{\mathrm{MT}}^{\mathrm{min}}$ of the considered system. 
	
	SR calculations serve for comparison with the other methods to investigate the magnitude of SOC effects. To determine the advantages of the SVLO over the SV method, we have carefully monitored the convergence of all considered quantities with respect to the number of SV(LO) basis functions. The NP method, as described in Ref.~\onlinecite{gulans_2022}, is used as a reference for this assessment. 
	
	For SR and SV calculations, we employ SR LOs. As we mainly address $p$ states in our examples, we label this case as $p$. The LO set including Dirac-type LOs, referred to as $p_{1/2}$, is constructed by adding to the SR LOs two $p_{1/2}$-type LOs for each $p$ state. Due to their $p$ character, each LO gives rise to three degenerate basis functions. The method is, however, fully general such to include relativistic LOs of other characters. For instance, we explore the effect of $d_{3/2}$ LOs in MoS$_2$ since its valence band maximum (VBM) and conduction band minimum (CBm) exhibit predominant $d$-character~\cite{Cappelluti_2013}. As the impact on the total energy and the electronic structure turns out to be negligible, however, we do not include this case in the following analysis. In the other materials, we additionally investigate the effects of $p_{3/2}$ LOs, by replacing the $p$ LOs. Due to their similar behavior near the nuclei, their impact is, however, only of the order of 10$^{-2}$ eV or smaller, which is within our convergence criteria. For this reason, we do not consider them further. $p_{1/2}$ LOs are used in SVLO and in the corresponding NP reference when specified. The number of LOs used in the different systems is displayed in Table~\ref{tab:systems_sizes} together with the size of the LAPW basis and the number of occupied valence states. To obtain the band-gap position for Bi$_{2}$Te$_{3}$, for the different sets of LOs ($p$-type and the $p_{1/2}$-type), we perform on top of the self-consistent NP calculation an additional iteration with a 48$\times$48$\times$48 $\bm{k}$-mesh. The so determined respective $\bm{k}$-points are then included in the band-structure path, from which we extract the final values of the energy gaps in the SV and SVLO calculations. 
	
	All input and output files are available at NOMAD \cite{Draxl_2019}.
	
	\begin{table}[h!]
		\centering
		\caption{Basis functions considered in the calculations of the five studied systems. $N_{\mathrm{LAPW}}$ is the number of LAPWs, $N_\text{LO}$ ($N_\text{LO}^{1/2}$) the number of LO basis functions for calculations without (with) Dirac-type LOs. The last column shows the number of occupied valence states, $N_{\mathrm{occ}}$. }
		\begin{tabular}{l|cccc}
			Material & $N_{\mathrm{LAPW}}$ & $N_\text{LO}$& $N_\text{LO}^{1/2}$ & $N_{\mathrm{occ}}$ \\
			\hline
			Xe &   138    &   26
			&38
			& 13   \\
			MoS$_2$ & 939      &  35      & -      &   13 \\
			PbI$_2$  &   318   & 73
			& 109
			&  33\\
			CsPbI$_3$  &   3236   &  496 
			&  736 & 228   \\
			Bi$_2$Te$_3$  &  895    &  141   &    201   &  57\\
		\end{tabular}
		\label{tab:systems_sizes}
	\end{table}
	
	\section{Results}
	
	\subsection{Xe}
\begin{figure}[ht!]
		\includegraphics[width=8.5cm]{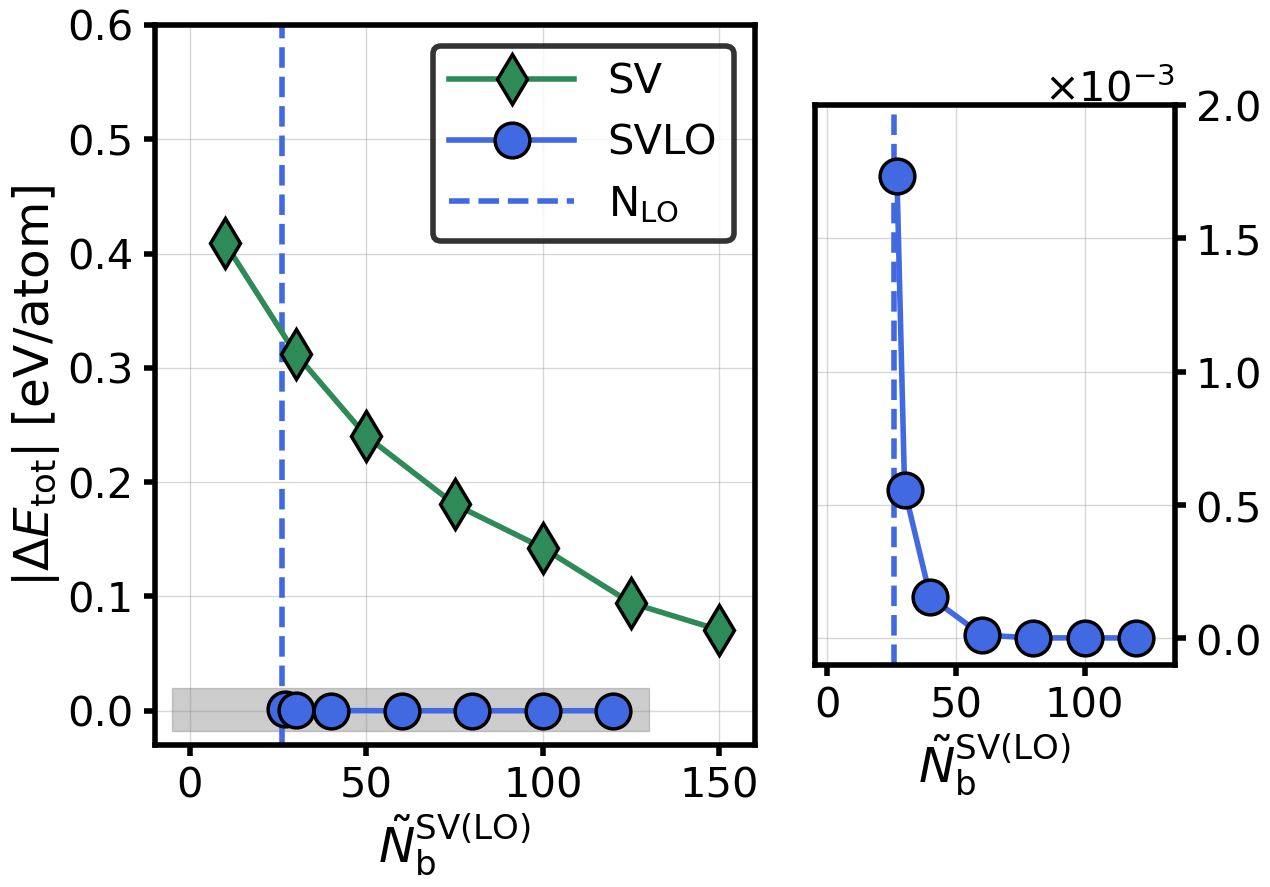}
		\caption{Convergence behavior of the total energy with respect to the number of basis functions $\tilde{N}_\mathrm{b}^{\mathrm{SV(LO)}}$ (excluding occupied states). The energy of the NP calculation is taken as a reference. Blue circles indicate the SVLO scheme, green diamonds the conventional SV treatment. The right panel zooms into the gray region where the SVLO method converges.}
		\label{fig:Tot_E_Xe}
	\end{figure}

	Our analysis starts with solid Xe. Fig.~\ref{fig:Tot_E_Xe} shows the convergence behavior of the total energy with respect to the number of basis functions, taking the NP calculation as a reference. For comparison, we also show the results of the conventional SV method. Since we employ the same number of occupied states in the SV and the SVLO method, and for the $p$ and $p_{1/2}$ sets of LOs (Table~\ref{tab:systems_sizes}), the number of basis functions on the x-axis does not include the occupied states, \ie $\tilde{N}_{\mathrm{b}}^{\mathrm{SV(LO)}}={N}_{\mathrm{b}}^{\mathrm{SV(LO)}}-N_{\mathrm{occ}}$. ${N}_{\mathrm{b}}^{\mathrm{SV(LO)}}$ is defined in Eq.~\ref{eqn:N_b_SV_LO} that applies to both methods. The number of LO basis functions in the equation is also predetermined, therefore, the increase in $\tilde{N}_\mathrm{b}^{\mathrm{SV(LO)}}$ reflects only the number of unoccupied bands. We will always refer to $\tilde{N}_\mathrm{b}^{\mathrm{SV(LO)}}$ when discussing the basis-set size. 
	
\begin{figure*}[ht!]
		\centering
		\includegraphics[width=0.9\textwidth]{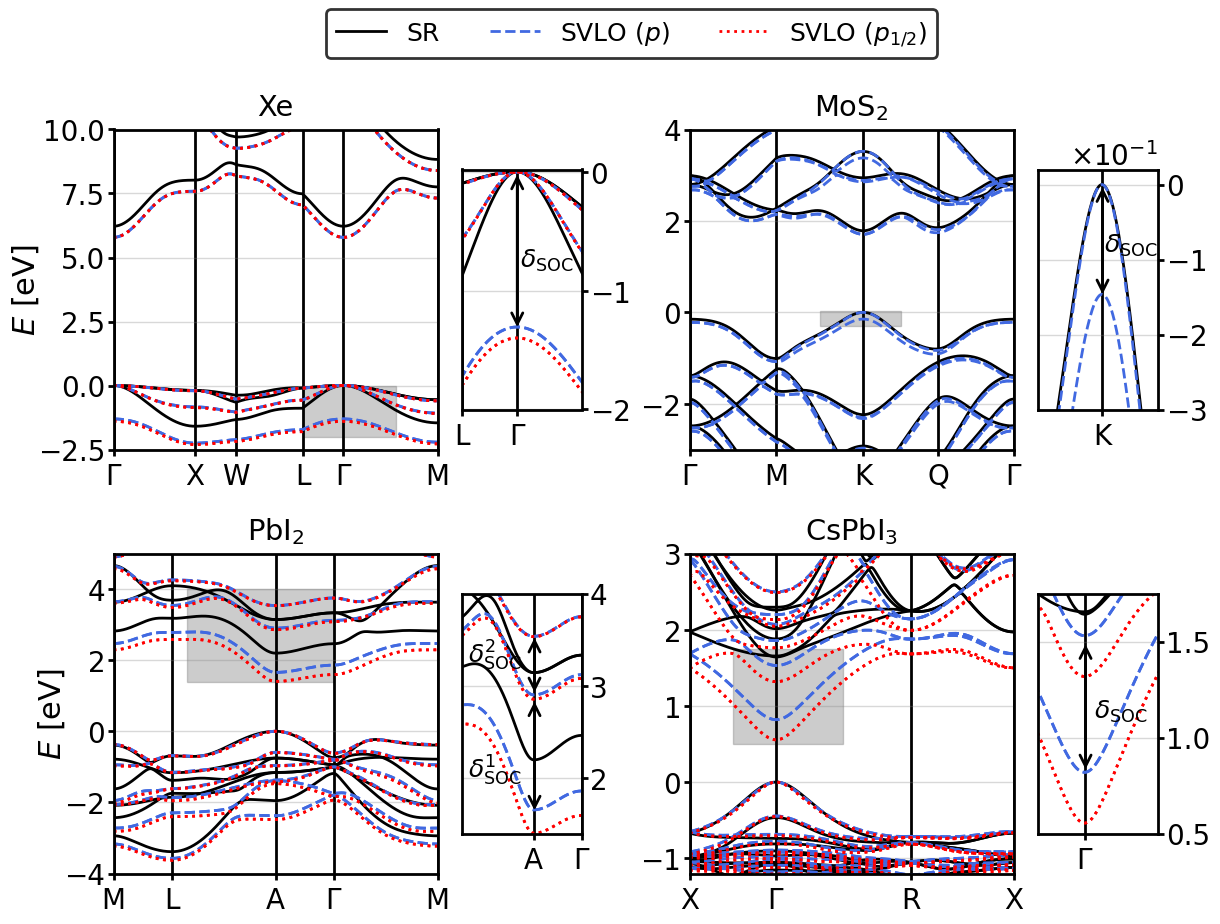}
		\caption{Band structures of Xe (upper-left panel), Mo$_2$ (upper-right panel), PbI$_2$ (lower-left panel), and CsPbI$_3$ (lower-right panel) computed with different methods and types of local orbitals. Black lines correspond to SR calculations and blue (red) lines to the SVLO method without (with) Dirac-type orbitals. The VBM is set to 0. At the right of each panel, we zoom into the corresponding region indicated by a gray box.}
		\label{fig:collection_bands}
	\end{figure*}

	In the case of Xe, we consider 26 SR LO basis functions (see Table~\ref{tab:systems_sizes}). 
	\begin{figure*}[ht!]
		\centering
		\includegraphics[width=0.8\textwidth]{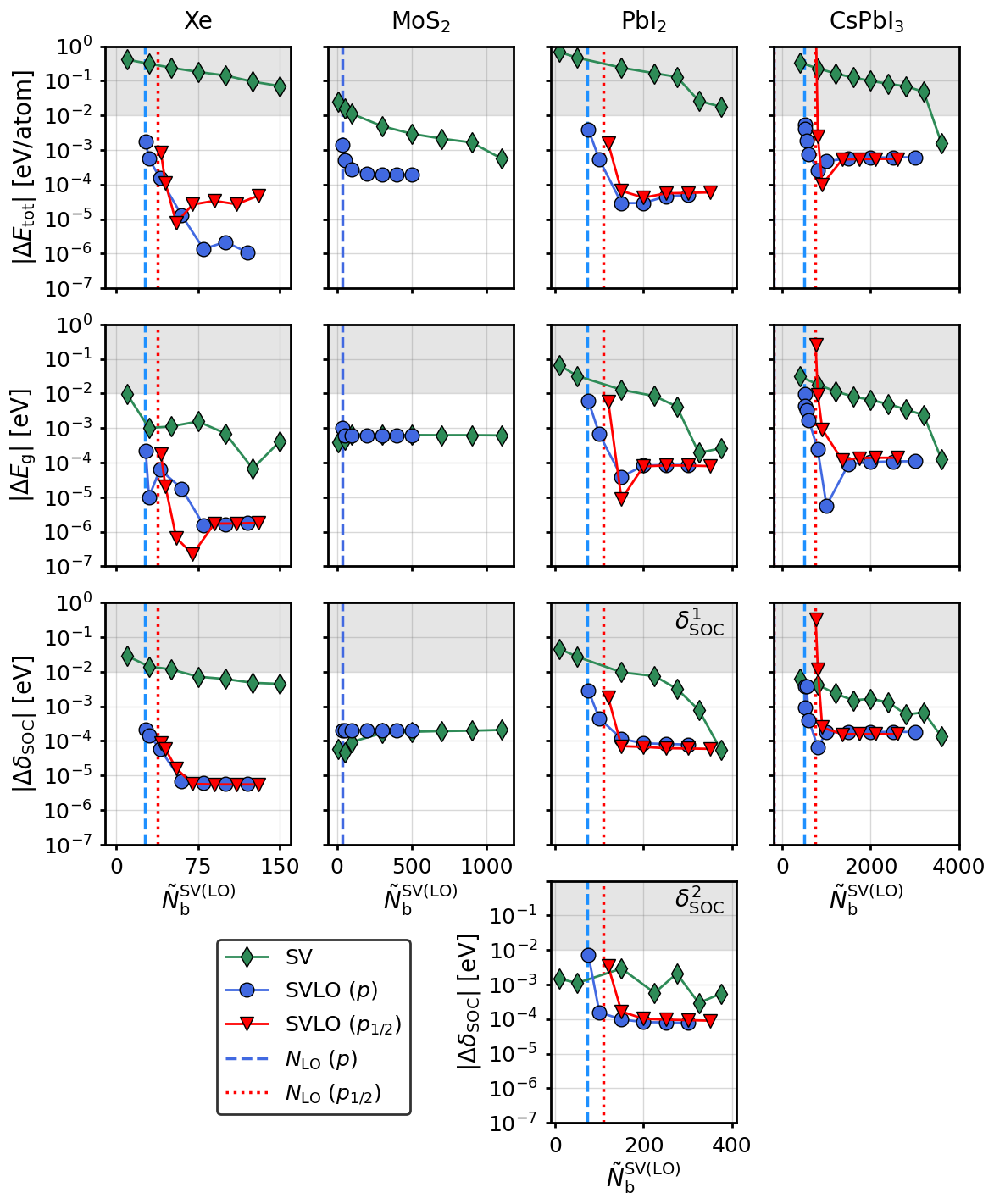}
		\caption{Convergence behavior of total energy, energy gap, and SOC splitting in Xe, MoS$_2$, PbI$_2$, and CsPbI$_3$, with respect to the number of basis functions used in the SV(LO) methods. Note the logarithmic scale on the y-axes. For the energy differences, the NP results serve as a reference. In the NP reference calculations, we employ sets of $p$ or $p_{1/2}$ LOs, depending on the method to compare with. Green diamonds stand for the SV method with SR LOs. All other results are obtained with the SVLO method, using different types of LOs: those obtained using SR (Dirac-type) orbitals are indicated by blue circles (red triangles). The vertical lines mark the respective number of LO basis functions. For PbI$_2$, we  display $\delta_{\mathrm{SOC}}^2$ and the energy difference $\delta_{\mathrm{SOC}}^1$ (both indicated in Fig.~\ref{fig:collection_bands}). The gray shaded areas are guides to the eye for highlighting the points which are within the target precision ($10^{-2}$ eV/atom for the total energy and $10^{-2}$ eV for the other quantities).}
		\label{fig:collection_Delta_E}
	\end{figure*}
	Strikingly, the total-energy differences obtained by the SVLO method stay within $2\times 10^{-3}$ eV/atom when employing a total number of basis functions comparable with the number of LO basis functions, while the SV method requires all available FV states to reach values even one order of magnitude larger ($7\times 10^{-2}$ eV/atom). To visualize this behavior better, Fig.~\ref{fig:collection_Delta_E} depicts the convergence of both methods on a logarithmic scale. We can observe that the SVLO method reaches convergence within $10^{-6}$ eV/atom with $\sim$80 basis functions. In this figure, we also analyze the convergence of the energy gap, $E_{\mathrm{g}}$, and the SOC splitting at the $\Gamma$ point, $\delta_{\mathrm{SOC}}$. When SOC is considered, $E_{\mathrm{g}}$ decreases by 0.43~eV due to the splitting of the (disregarding spin) three-fold degenerate VBM into a single state and a double-degenerate state by about 1.30 eV (see also Table \ref{tab:Egap} and Fig. \ref{fig:collection_bands}). For both quantities, we observe that the SVLO method reaches a precision of the order of $10^{-4}$ eV already with a number of basis functions comparable with the number of LO functions; with approximately 80 basis functions even two orders of magnitude better. In contrast, the SV treatment, employing all available FV KS eigenstates, only converges within $10^{-3}$ eV  and $10^{-2}$ eV for the energy gap and the SOC splitting, respectively. If we consider a target precision often used for production calculations such as 10$^{-2}$ eV/atom for the total energy and 10$^{-2}$ eV for energy gaps and SOC splittings, the advantage of the SVLO method is particularly considerable for the total energy. In contrast, the SV energy gap reaches the target precision at a number of empty states smaller than the number of LO basis functions, and the corresponding SOC splitting requires approximately 75 empty states.
	\begin{table*}[ht!]
		\caption{Energy gaps, $E_{\mathrm g}$, and SOC splittings, $\delta_{\mathrm{SOC}}$, of the considered materials computed with the SVLO method for different sets of LOs. For comparison, scalar-relativistic (SR) results are shown. For Bi$_2$Te$_3$, that does not exhibit any SOC splitting, we show the energy difference between the highest valence band (VB) and the lowest conduction band (CB) at $\Gamma$, $E_{\Gamma\rightarrow\Gamma}$. Note that in this material, SOC not only changes the magnitude of the gap but also the position of the VBM and the CBm. Both are again altered when Dirac-type LOs are considered.}
		\begin{tabular}{l|ccccc|ccccc|c}
			\multirow{2}{*}{Method}    & \multicolumn{5}{c|}{$E_{\mathrm{g}}$ [eV]} & \multicolumn{5}{c|}{$\delta_{\mathrm{SOC}}$ [eV]} & $E_{\Gamma\rightarrow\Gamma}$ [eV]\\
			\cline{2-12}
			& Xe & MoS$_2$ &PbI$_2$ &  CsPbI$_3$ & Bi$_2$Te$_3$& Xe & MoS$_2$ & PbI$_2$ ($\delta_{\mathrm{SOC}}^1$) & PbI$_2$ ($\delta_{\mathrm{SOC}}^2$)  & CsPbI$_3$ & Bi$_2$Te$_3$ \\
			\hline
			SR & 6.22     &  1.78  &  2.20   & 1.64    & 0.25 ($\Gamma\rightarrow\Gamma$) & -& -& 0.94 &-& - & 0.25 \\
			SVLO ($p$) &   5.79   & 1.71   &  1.66 
			& 0.82 
			& 0.10 (B$\rightarrow $B) &   1.30
			&  0.15  & 1.25   
			& 0.63
			& 0.71 
			& 0.58\\
			SVLO ($p_{1/2})$ & 5.79  
			&   &  1.40 
			&  0.55       & 0.03 (D$\rightarrow $C) & 1.40   &  &  1.45  &    0.68
			&  0.76   & 0.69       \\
		\end{tabular}
		
		\label{tab:Egap}
	\end{table*}
	
	Dirac-type LOs turn out to be significant for the SOC splitting which increases by $0.1$ eV (Table~\ref{tab:Egap}) upon adding four $p_{1/2}$-type LOs, each of them contributing three degenerate basis functions (Table~\ref{tab:systems_sizes}). Their effect on the energy gap is negligible. The convergence behavior of the energy gap and the SOC splitting with respect to the number of basis functions is comparable to that of the SVLO method with SR LOs (Fig.~\ref{fig:collection_Delta_E}). Contrarily, the total energy converges to a worse precision (within $10^{-4}$ eV/atom). Also with Dirac-type LOs the analyzed quantities reach the targeted precision with a few empty states in addition to the LO basis functions. In the Appendix, we explain why the SV and the SVLO method do not converge to the same precision.
	
	\subsection{MoS$_2$}
	The transition-metal dichalcogenide MoS$_2$ is among the most studied 2D materials, and a candidate for many applications in optoelectronics. SOC reduces the energy gap by 0.07 eV only (Table~\ref{tab:Egap}), caused by a splitting of the VBM. Although this splitting is rather small, \ie 0.15 eV, it is fundamental as, being not considered, could lead to the unphysical prediction of an indirect band gap~\cite{Pisarra_2021}. Moreover, the splitting at the $\mathrm{K}$-point of the Brillouin zone (BZ) is essential for the accurate description of the optical spectra ~\cite{Diana_2016,Marsili_2021}. Regarding the convergence behavior (second column of Fig.~\ref{fig:collection_Delta_E}), we observe a small improvement of the SVLO method over the SV method for the total energy: With around 40 basis functions, the SVLO (SV) method reaches a precision of the order of $10^{-3}$  eV/atom (10$^{-2}$  eV/atom). For the energy gap and the SOC splitting, both methods reach convergence with a few basis functions and reproduce the NP treatment with a precision of the order of $10^{-4}$ eV.
	
	\subsection{PbI$_2$}
	Lead iodide, PbI$_2$, is a semiconductor used for detectors, and it is also a precursor for the heavily investigated solar-cell materials, the lead-based halide perovskites. Like in the latter, in PbI$_2$, the SOC effects are massive. 
	The band gap reduces by 0.54 eV (Table~\ref{tab:Egap}, Fig.~\ref{fig:collection_bands}) and (disregarding spin) the two-fold degenerate second conduction band (CB) experiences a splitting of $\delta_{\mathrm{SOC}}^{2}$= 0.63 eV. There is also an increase in the energy distance between the CBm and the second CB which is 0.31 eV (Table~\ref{tab:Egap}). For convenience, we label this energy difference as $\delta_{\mathrm{SOC}}^{1}$. For PbI$_2$, the advantages of the SVLO method over SV are considerable. With a number of basis functions comparable to the number of LO functions (here 73, see Table~\ref{tab:systems_sizes}), the SVLO method has reached the target precision for all considered quantities (Fig.~\ref{fig:collection_Delta_E}). Contrarily, the SV method, requires basically all empty states ($\sim$375 basis functions) for reaching the target precision for the total energy; $\sim$225 empty bands are needed for the energy gap and $\sim$150 for $\delta_{SOC}^{1}$, while only $\sim$10 empty states for $\delta_{SOC}^{2}$. Except for the total energy, for which the SV method converges to a precision of the order of 10$^{-2}$ eV/atom and the SVLO method of the order of 10$^{-5}$ eV/atom, both approaches converge to comparable precision. 
	
	For an accurate prediction of the electronic structure, $p_{1/2}$-type LOs are crucial (Fig.~\ref{fig:collection_bands}). We add 4 for each species, with a total of 36 LOs basis functions (see Table~\ref{tab:systems_sizes}). They reduce the energy gap further by 0.26 eV and increase $\delta_{\mathrm{SOC}}^1$ by additional 0.20 eV. $\delta_{\mathrm{SOC}}^2$ increases by 0.05 eV only (Table~\ref{tab:Egap}). The convergence behavior with $p_{1/2}$-type LOs is overall comparable to that with SR LOs. Note that the two curves appear shifted by these 36 additional basis functions. Although this number of LO basis functions is considerable for such a system (see Table~\ref{tab:systems_sizes}), the speed-up with respect to the SV method is significant also when Dirac-type LOs are employed.     
	
	\subsection{CsPbI$_3$}
	CsPbI$_3$ is among the most studied inorganic metal halide perovskites \cite{Zhun_2021,Vona_2022}. We consider it in the orthorhombic $\gamma$-phase that contains 20 atoms. Being composed by three heavy elements, SOC effects are enormous. The band gap decreases from 1.64 eV with SR to 0.82 eV when SOC is considered (Table~\ref{tab:Egap}). This is caused by a 0.71 eV splitting of the (disregarding spin) two-fold degenerate CBm (Fig.~\ref{fig:collection_bands}). When Dirac-type LOs are added, the gap further reduces by 0.27 eV, while the splitting increases by only 0.05 eV (Table~\ref{tab:Egap}). 
	
	Although SV and SVLO($p$) converge to the same results within the target precision, the computational effort required for the two approaches is noticeably different. In the limit of large unit cells --CsPbI$_3$ is the largest one considered here-- the dominant contribution to the run time comes from the tasks that scale cubically with respect to the system size. These tasks include the construction of the Hamiltonian matrices and the diagonalization. As shown in Fig.~\ref{fig:collection_Delta_E}, a converged SV calculation requires that essentially all unoccupied bands are included for solving the full problem. In this light, SV does not offer any advantage over the NP approach. In contrast, to converge the SVLO($p$) calculation, it is sufficient to use a significantly smaller basis with $N_\mathrm{occ}=228$, $N_\mathrm{LO}=496$, and $N_\mathrm{unocc}\sim 0$ (see Table~\ref{tab:systems_sizes}). Taking into account the spin degrees of freedom, the size of the Hamiltonian matrix in the SV step is $\sim$1500. As discussed above, diagonalization is also required in the FV step, where the dimension of the SR Hamiltonian is $\sim$3800. This step is therefore the most computationally intensive in this example. Compared to the NP calculation, we find that total time spent on the FV and SV steps is reduced by a factor of 3.6. Finally, the inclusion of $p_{1/2}$-type LOs increases $N_\mathrm{LO}$ to 736 and thus also slightly increases the size of the SV diagonalization problem.
	
	\subsection{Bi$_2$Te$_3$}
\begin{figure}[h!]
		\centering
		\includegraphics[width=0.9\columnwidth]{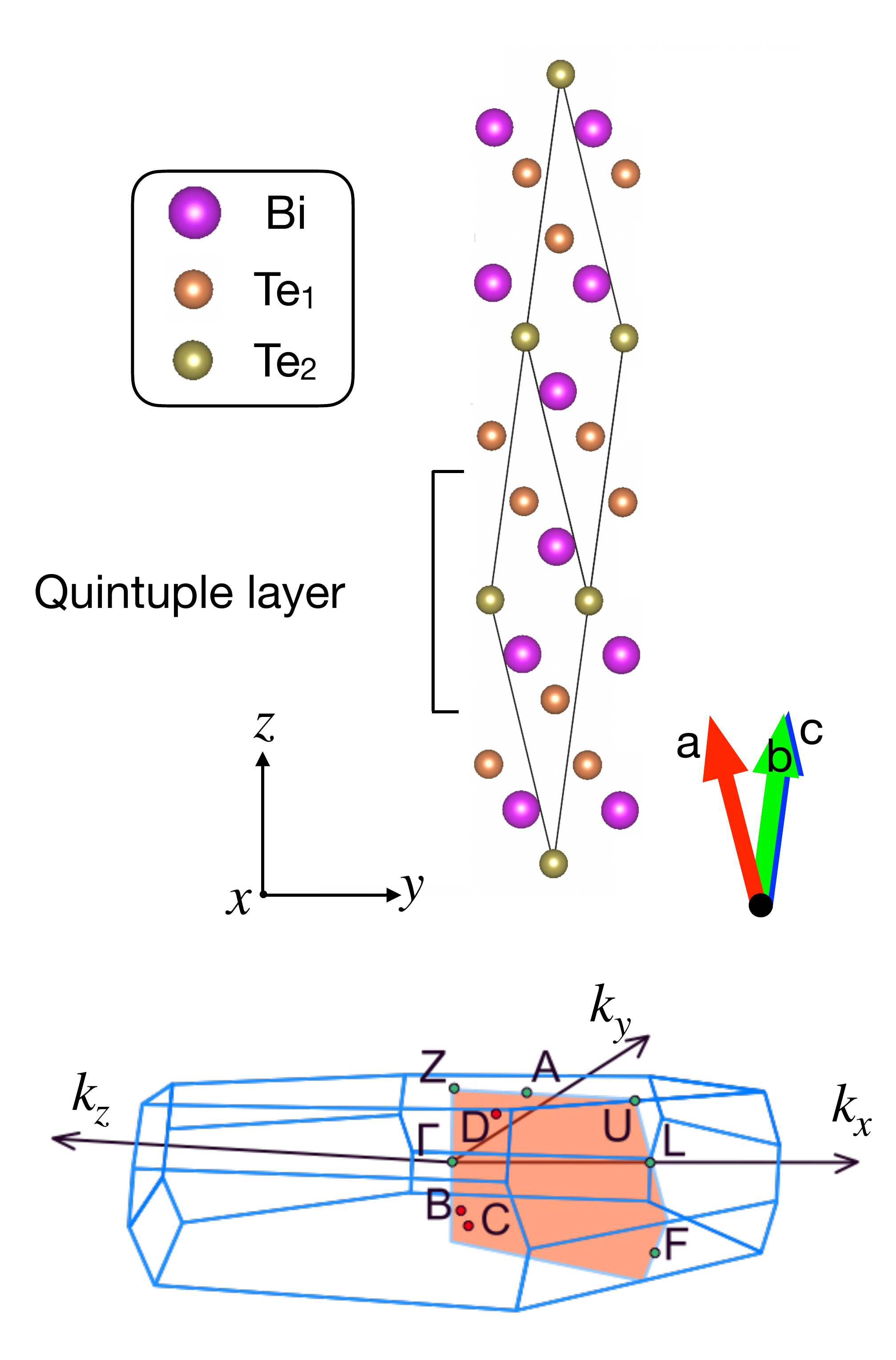}
		\caption{Top: Crystal structure of Bi$_2$Te$_3$, built by Bi (pink) and two chemically inequivalent Te atoms (Te$_1$ orange, Te$_2$ gold). Bottom: Corresponding Brillouin zone. The mirror plane containing the points depicted in the band structure in Fig.~\ref{fig:Bi2Te3_band} is indicated in red.}
		\label{fig:Bi2Te3_structure}
	\end{figure}

	Bi$_2$Te$_3$ is a topological insulator with a single Dirac cone at $\Gamma$ \cite{Zhang_2009,Chen_2009}. It is characterized by strong SOC effects, shifting the fundamental band gap from $\Gamma$ to an off-symmetry point in the mirror plane of the first Brillouin zone that is displayed in the bottom panel of Fig.~\ref{fig:Bi2Te3_structure}. The positions of the VBM and CBm are highly sensitive to the structure and the choice of the exchange-correlation functional, thus there are controversial results present in the literature. Ref.~\onlinecite{Fang_2019} presents an overview of this diversity that increases when more accurate methods, such as the $GW$ approximation, are applied ~\cite{Emmanouil_2010,Aguilera_2013}. All these aspects together make Bi$_2$Te$_3$ computationally challenging.
	
	Bi$_2$Te$_3$ crystallizes in a rhombohedral structure with R-3m symmetry, shown in the top panel of Fig.~\ref{fig:Bi2Te3_structure}. It consists of five layers, with alternating Te and Bi sheets, repeated along the z-direction. There are two chemically inequivalent Te sites.
	
\begin{figure}[ht!]
		\centering
		\includegraphics[width=0.9\columnwidth]{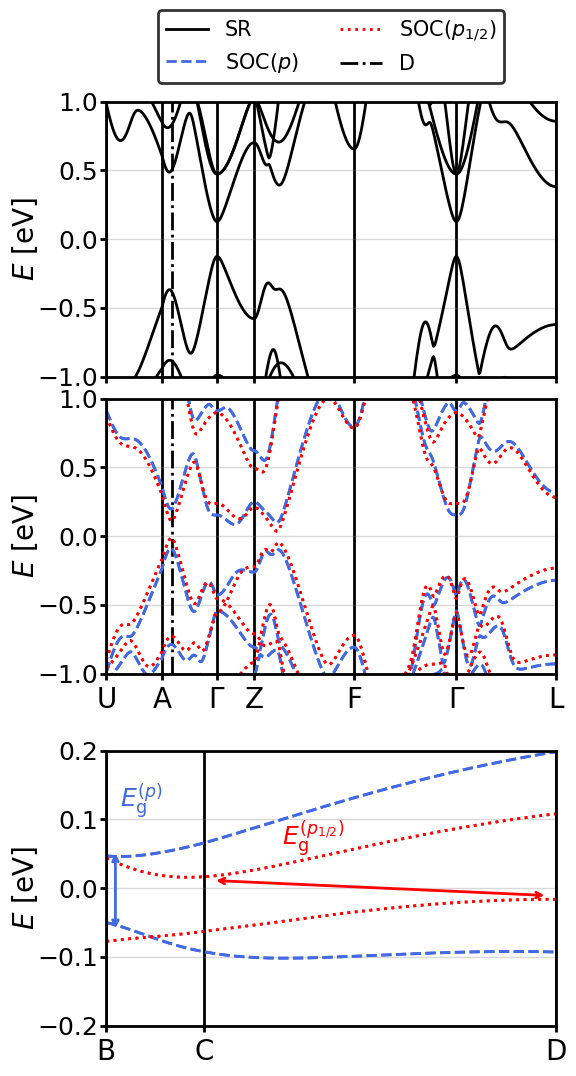}
		\caption{Band structure of Bi$_2$Te$_3$, computed without (top panel) and with SOC (other panels). The  coordinates of the high-symmetry points are U~(0.823,0.339, 0.339), Z~(0.5,0.5,0.5), F~(0.5,0.5,0.0), L~(0.5,0.0,0.0); those of points A, B, C, and D are given in the text. The dashed vertical lines in the two top panels indicate the position of point D. The bottom panel zooms into the region of the band edges, showing the direct (indirect) band gap computed with $p$ ($p_{1/2}$) LOs. Note that, differently from Fig.~\ref{fig:collection_bands}, the energy zero is not at the VBM but in the middle of the band gap.} 
		\label{fig:Bi2Te3_band}
	\end{figure}

	Including SOC, the band structure undergoes significant changes that are further enhanced when Dirac-type LOs are added~\cite{Larson_2003,Huang_2008} (top and middle panels of Fig.~\ref{fig:Bi2Te3_band}). A relevant difference is observed at $\Gamma$ where the valence band (VB) and the CB obtained from SOC calculations show a hump as a consequence of the band-inversion characteristic of this material~\cite{Zhang_2009,Fang_2019}. Differently from similar topological insulators, the hump is well preserved in spinor $GW$ calculations, which include the off-diagonal elements of the self-energy, even though the band dispersion is strongly altered~\cite{Aguilera_2013}. 
	SR calculations lead to a direct band gap of 0.25 eV at $\Gamma$ (Table~\ref{tab:Egap}). By adding SOC --but no Dirac-type LOs-- it reduces to 0.10 eV and is located at point B=(0.67, 0.58, 0.58), which appears sixfold in the BZ. Our results are comparable with those of Ref.~\onlinecite{Huang_2008} and Ref.~\onlinecite{Scheidemantel_2003}. In the former, a direct band gap of 0.13~eV at (0.667,0.571, 0.571) was measured, while in the latter, a value of 0.11 eV was computed, but different from our result, it was reported to be indirect. However, VBM and CBm are very close to each other being located at (0.652, 0.579, 0.579) and (0.663, 0.568, 0.568), respectively. One may assign these differences to the use of different $\bm{k}$-grids and crystal structures (here we use $a=10.44$ \AA~ and $\theta=24.27^{\circ}$~\cite{Akshay_2017}, while Refs.~\cite{Scheidemantel_2003,Huang_2008} use an experimental structure with $a=10.48$ \AA~ and $\theta=24.16^{\circ}$). 
	
	By adding 4 $p_{1/2}$-type LOs for each species, \ie a total of 60 basis functions (Table~\ref{tab:systems_sizes}), the gap reduces to 0.03 eV and becomes indirect. In the bottom panel of Fig.~\ref{fig:Bi2Te3_band}, we observe that the VB is lowered at B and raised at D=(0.52, 0.35, 0.35) where the VBM is now located. The CB is not altered at B but lowered at C=(0.65, 0.54, 0.54) which is the approximate location of the CBm (the resolution being limited by the $48\times48\times48\; \bm{k}$-mesh). Points C an D are six-fold degenerate. D is located between $\Gamma$ and A=(0.64, 0.43, 0.43), which lies on the path between $Z$ and $U$. C and B are close to $Z\rightarrow F$. Larson \cite{Larson_2003} and Huang and coworkers \cite{Huang_2008} obtained gaps of 0.05~eV and 0.07~eV, respectively, with $p_{1/2}$-type LOs. The locations of the band extrema slightly differ between the three works where ours is in better agreement with that of Larson~\cite{Larson_2003}. 
	
\begin{figure*}[hb!]
		\centering
		\includegraphics[width=0.9\textwidth]{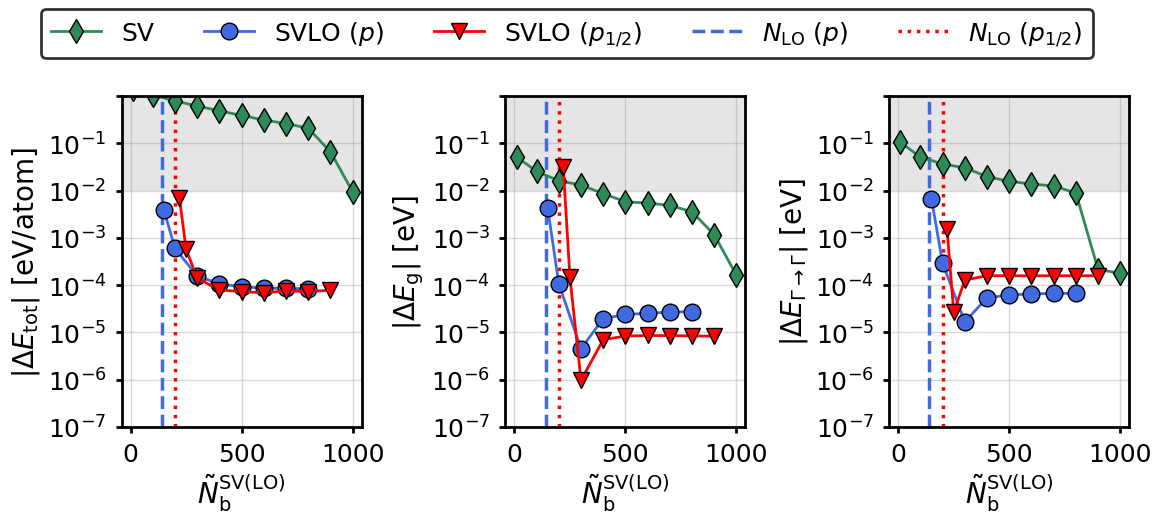}
		\caption{Convergence behavior of total energy, energy gap, and energy difference between the highest VB and the lowest CB at $\Gamma$ with respect to the number of second-variational basis functions, $\tilde{N}_{\mathrm{b}}^{\mathrm{SV(LO)}}$, for Bi$_2$Te$_3$.}
		\label{fig:Bi2Te3_conv}
	\end{figure*}

	As evident from Fig.~\ref{fig:Bi2Te3_conv}, for Bi$_2$Te$_3$, our new method has clear advantages over the conventional SV method, which reaches the target precision for the total energy only with basically all available FV KS states (about $\sim$1000, see Table~\ref{tab:systems_sizes}), while the SVLO method requires a basis-set size comparable with the number of LO basis functions (141 for the $p$-set and 201 for the $p_{1/2}$-set). The SVLO method converges in either case within a precision of $10^{-4}$ eV/atom. Like for the other materials, the electronic structure obtained by SV, converges faster than the total energy, but the convergence is still not comparable with that of the SVLO method. To obtain an energy gap within a precision of $10^{-2}$ eV, the SV method requires about twice the number of basis functions (without including the occupied states); to obtain $E_{\Gamma\rightarrow\Gamma}$ with a precision of $\sim 10^{-4}$ eV, the basis size needs to be further doubled. The band gap converges to a precision of $\sim 10^{-5}$ eV, while the SV method cannot go lower than $10^{-4}$ eV. 
	
	\section{Discussion and conclusions}
	In this work, we have introduced a novel approach --the SVLO method-- to treat spin-orbit coupling in DFT calculations efficiently. It allows us to obtain rapid convergence and highly precise results, \eg band energies within the order of 10$^{-4}$ eV or even better. SOC splittings and total energies within a precision of 10$^{-2}$ eV and 10$^{-2}$ eV/atom, respectively, can actually be obtained with a number of basis functions that is comparable to the number of occupied states plus a set of LOs. Its efficiency is owing to the fact that SOC effects mainly come from regions around the atomic nuclei where atomic-like functions play the major role in describing them. We have demonstrated this method with examples of very different materials. The use of the SVLO method is most efficient when SOC effects are strong. In the cases, we also observe significant contributions of $p_{1/2}$ LOs. Obviously, the overall gain of our method is getting more pronounced the bigger the system is. In summary, by providing a method that allows for reliable and efficient calculations of SOC, our work contributes to obtaining highly-accurate electronic properties at the DFT level. 
	
	\begin{figure*}[ht!]
		\centering
		\includegraphics[width=0.9\textwidth]{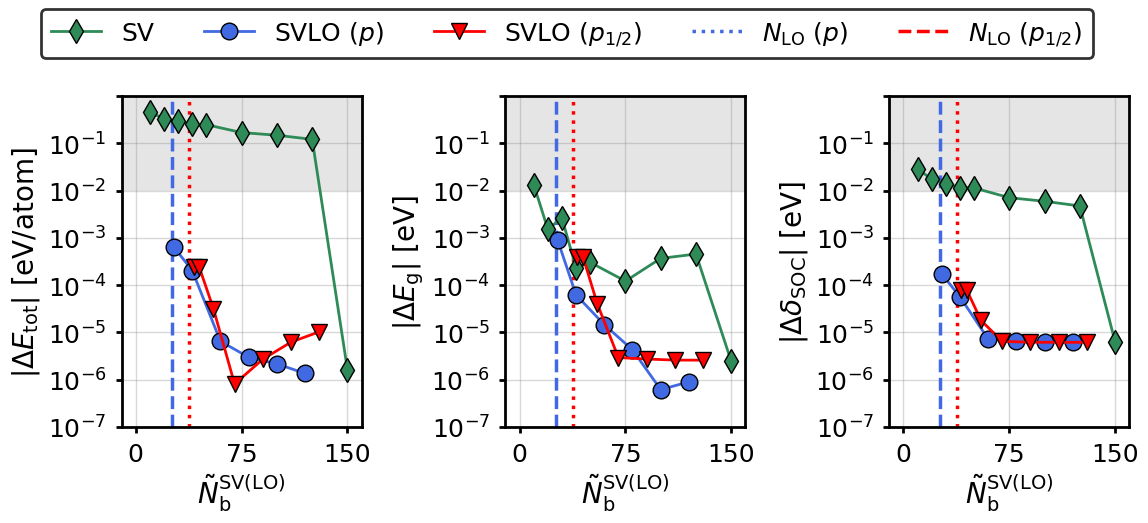}
		\caption{Same as Fig.~\ref{fig:collection_Delta_E} for Xe, but for one $\bm{k}$-point only.} 
		\label{fig:Xe_1k_point}
	\end{figure*}
	
	\acknowledgements
	This work was supported by the German Research Foundation within the priority program SPP2196, Perovskite Semiconductors (project 424709454) and the CRC HIOS (project 182087777, B11). A.G. acknowledges funding provided by European Regional Development Fund via the Central Finance and Contracting Agency of Republic of Latvia under the grant agreement 1.1.1.5/21/A/004. Partial support from the European Union’s Horizon 2020 research and innovation program under the grant agreement N° 951786 (NOMAD CoE) is appreciated.
	\\ \\
	\appendix
	\section*{Appendix}
	In the examples discussed above, the SV method often converges to worse precision than the SVLO method. This may appear counter intuitive since the two methods should be equivalent if the SV basis includes all available FV KS states. The reason for the seeming discrepancy comes from the fact that the LAPW basis-set size may be different at different $\bm{k}$-points, depending on their symmetry. In contrast, in the SV method, the size of the basis is controlled by an input parameter and limited by the number of available FV KS orbitals. In our implementation, the same number is considered for all $\bm{k}$-points. In Fig.~\ref{fig:Xe_1k_point}, we show for the example of Xe that --when carrying out the SVLO calculation with a single $\bm{k}$-point-- the two methods reach the same precision (of the order of $10^{-6}$ eV) for all analyzed properties. In this case, all KS orbitals can be used as basis functions in the SV method. We emphasize, however, that the inclusion of all KS states is not efficient and thus not desirable anyway. 
	
	
\end{document}